\documentclass[noshowpacs,prl,aps,superscriptaddress,floatfix,twocolumn,footinbib]{revtex4-2} 
\usepackage[utf8]{inputenc}
\usepackage{lineno,mathtools,url}
\usepackage{amsmath}
\usepackage{soul}
\usepackage{amssymb,mathrsfs}
\usepackage{physics}
\usepackage{graphicx}
\usepackage{hyperref}

\hypersetup{
     colorlinks=true,
     linkcolor=blue,
     filecolor=blue,
     citecolor = orange,      
   }
\usepackage[dvipsnames]{xcolor}
\usepackage{dsfont}
\usepackage{mathbbol} 
\usepackage[normalem]{ulem}
\allowdisplaybreaks
\begin{document}

\title{
Sound, Superfluidity and Layer Compressibility in a Ring Dipolar Supersolid
}

\author{Marija \v{S}indik}\email[]{These 2 authors contributed equally to the paper}
\author{Tomasz Zawi\'slak}\email[]{These 2 authors contributed equally to the paper}
\author{Alessio Recati}\email[]{Corresponding Author: alessio.recati@ino.cnr.it}
\author{Sandro Stringari}
\affiliation{Pitaevskii BEC Center, CNR-INO and Dipartimento di Fisica, Universit\`a di Trento, Via Sommarive 14, 38123 Povo, Trento, Italy}

\begin{abstract}
We propose a protocol to excite the Goldstone modes of a supersolid dipolar Bose-Einstein condensed gas confined in a ring geometry. 
By abruptly removing an applied periodic modulation proportional to cos($\varphi$), where $\varphi$ is the azimuthal angle, we explore the resulting oscillations of the gas, by solving the extended Gross-Pitaevskii equation. The value of the two longitudinal sound velocities exhibited in the supersolid phase are analyzed using the hydrodynamic theory of supersolids at zero temperature, which explicitly takes into account both the superfluid and the crystal nature of the system. This approach allows for the determination of the layer compressibility modulus as well as of the superfluid fraction $f_S$, in agreement with the Leggett estimate of the nonclassical moment of inertia. 

\end{abstract}

\maketitle

A key consequence of the  spontaneous breaking of continuous symmetries is the occurrence of Goldstone modes, which, in the presence of finite range interactions,  take the form  of gapless excitations in the long wavelength limit. The identification and the experimental observation of the Goldstone modes, then, represents a question of central interest in various fields of science, including  elementary particle physics, magnetism, superfluidity, and superconductivity.  The recent realization of supersolidity has raised the question of the identification of the corresponding Goldstone modes which are the consequence of the spontaneous and simultaneous breaking of phase symmetry and translational invariance, ensuring the nonintuitive coexistence of superfluid and crystal features. From the theoretical side the study of the Goldstone modes in supersolids has a long history,  starting from the pioneering work of Andreev and Lifshitz~\cite{AndreevLifshitz} (see also ~\cite{Saalow,Son,Hydro-Rica2007,Dorsey2010,Hofmann2021}),  and more recent papers based on numerical simulations on  atomic Bose gases interacting with soft-core potentials~\cite{Boninsegni_SS2012,Macri_SS,Ancilotto2013}, spin orbit coupled gases (see the recent reviews~\cite{ARCMP,MartoneReview} and reference therein), condensation in multimode cavity \cite{natalia2023}, and dipolar gases~\cite{Roccuzzo1} (see also the recent perspective~\cite{NaturePersp} and references therein). First experimental evidence for the occurrence of dispersive Goldstone modes in a supersolid has been recently reported in the case of a dipolar gas confined in a harmonic trap, where the modes take the form of discretized oscillations and in particular with the emergence of novel crystal-like oscillations as soon as one enters the supersolid phase~\cite{F2,S2,I2}. 

The use of harmonic trapping potentials, inducing the nonhomogeneity of the gas, together with the appearance of a  number of droplets that form the nonsuperfluid (crystal) component of the gas,  limits, however, the possibility to fully
appreciate the rich dynamics of the dipolar gas as a bulk
supersolid.
Theoretically,  the speeds of sounds for soft-core~\cite{Boninsegni_SS2012,Macri_SS} and dipolar~\cite{Roccuzzo1,Blakie2023sounds}  gases in the thermodynamic limit have already been the object of  numerical  simulations in the supersolid phase, based on the use of  proper  periodic boundary conditions. In particular, in the very recent work~\cite{Blakie2023sounds}, which appeared while completing our work, a detailed analysis   of the two longitudinal sounds along an infinite tube has been reported as  a function of the relevant interaction parameters. 
Experimentally, reaching the thermodynamic limit  in a dipolar gas is, however, strongly inhibited if one uses box potentials because of  the tendency of the dipoles to accumulate near the borders  of the wall, giving rise to typical edge effects~\cite{Lu-preEdge2010,roccuzzoEdge}.

For the above reasons, in this Letter we propose to use a ring potential, which naturally fulfills  the  periodic  boundary conditions  along the azimuthal angle direction, allowing for an efficient way to approach the thermodynamic limit.  Ring traps provide a simple realization of matter wave circuits, with important
perspectives in the emerging field of atomtronics \cite{RoadmapAtomTronic,Amico22}.
The longitudinal Goldstone modes of the ring trapped dipolar gas are  then excited by suddenly removing  a periodic annular perturbation    
and the resulting values of the sound velocities
are analyzed  using the hydrodynamic theory of supersolids. Supersolid hydrodynamics actually provides fundamental new insights on the origin of the two sound modes. In particular, employing the calculated values of the sound velocities,  we determine for the first time the layer
compressibility modulus and confirm the prediction for  the superfluid fraction of a
dipolar gas in the supersolid phase based on the nonclassical moment of inertia and its concurrence with Leggett's estimation~\cite{Leggett1998}.

\paragraph{The model and its quantum phases.}

In a dilute dipolar Bose gas, the atoms interact by a delta-contact potential $V_c({\bf r})=g\delta({\bf r})$, with the coupling constant $g=4\pi\hbar^2a/m>0$ fixed by the atomic mass $m$ and the s-wave scattering length $a$; and by the dipolar potential $V_{dd}({\bf r}) =\frac{\mu_0\mu^2}{4\pi}\frac{1-3\cos^2\theta}{|{\bf r}|^3}$  with $\theta$ the angle between ${\bf r}$ and the direction $z$ of the externally applied magnetic field, which aligns the atomic magnetic dipole moments $\mu$. The most important parameter to determine the zero temperature phase diagram of the gas is the ratio between the strengths of the contact and the dipolar interactions, $\epsilon_{dd} =a_{dd}/a$ with $a_{dd}=\mu_0\mu^2/12\pi\hbar^2$ the so-called dipolar length. For small enough $\epsilon_{dd}$ the system forms a Bose-Einstein condensate (BEC), while by increasing it beyond a certain threshold, in three-dimensional uniform configurations, the system collapses due to the attractive nature of the dipolar interaction. Confining the gas along the $z$ direction prevents this collapse, and three distinct phases occur: (i) a homogeneous BEC (superfluid phase), (ii) a supersolid phase
in a very small interval of $\epsilon_{dd}$, and (iii) a droplet crystal phase, i.e., independent droplets arranged in a crystal structure. In the present Letter we consider moreover that the gas is confined in the $x$-$y$ plane by a ring-shaped potential $V_\mathrm{ext}(r_\perp,z)=m\left[\omega_\perp^2(r_\perp-R_0)^2 + \omega^2_zz^2\right]/2$ with $r_\perp=\sqrt{x^2+y^2}$, of radius $R_0=7.64\,\mu$m and trap frequencies $\omega_z=\omega_\perp= 2\pi \cdot 100$ Hz, leading to the three phases reported in Fig.~\ref{fig:2}, calculated for $N=80000$ $^{164}$Dy atoms, corresponding to $a_{dd}=132 a_0$, where $a_0$ is the Bohr radius. 
We obtain a ring-shaped cloud of length $L\approx 49\,\mu$m~\footnote{The cloud's shape is influenced by long-ranged dipolar repulsion effectively increasing its radius.} and width $\text{\small FWHM}_{XY}$ changing from  $1.41\,\mu$m in the superfluid to $0.7\,\mu$m approaching the crystal phase (see Fig.~\ref{fig:2}). Because of magnetostriction, the cloud is elongated in the third direction with $\text{\small FWHM}_Z \approx 4\,\mu$m.
The system is numerically studied within the so-called extended Gross-Pitaevskii equation~\cite{Wachter2016} which in the last few years has been systematically employed to describe the equilibrium and dynamic properties of dipolar supersolids, in reasonably good agreement with the experimental findings.


\paragraph{The protocol.} We first determine the state of the gas by applying a small static perturbation of the form $-V_0\cos\varphi$, where $\varphi$ is the azimuthal angle along the ring, 
which produces stationary density modulations. We then suddenly set $V_0=0$, resulting 
in the excitation of the longitudinal phonon modes propagating along the ring.  Similar protocols have been already applied  to investigate the Doppler effect due to the presence of quantized vortices in a ring~\cite{Kumar2016} and, more recently, to investigate  the effect of superfluidity on the propagation of sound in a dilute Bose gas confined in a box in the presence of an external periodic potential~\cite{Chauveau2023}.   

An easy analysis of the response of the system can be obtained considering  sufficiently large ring sizes for which the ring can be mapped in a linear tube 
configuration with imposed periodic boundary conditions.  In particular, we assume that the length $L$ of the ring is much larger than its width, so that  one can safely identify $\cos(\varphi)$ with $\cos(qx)$, where  $q=2\pi/L$ is the wave vector of the longitudinal excitation and the variable $x$, with $0\le x \le L$, is the longitudinal coordinate along the tube. 
According to linear response theory, 
the quantity $F(t) =\langle \cos{\varphi}\rangle(t)$ should show, in the supersolid phase, a beating of two modes (see inset in Fig.~\ref{fig:4})
\begin{equation}\label{eq:df}
F(t)=V_0\sum_{i=\pm}\chi_i(q)\cos(\omega_i(q)t) 
\end{equation}
with $\omega_i(q)$ approaching, for sufficiently small $q$ (and hence large $L$), the linear phonon dispersion ${\omega_i(q)\simeq c_iq}$, with $c_+$ and $c_-$  hereafter called upper and lower sound velocities, respectively. The quantities $\chi_i(q)$, $i=\pm$,
define the contributions of the two modes to the static response and hence to the compressibility sum rule according to
\begin{equation}
\chi(q)=\chi_+(q)+\chi_-(q) =  \int_0^\infty d\omega\, \frac{S(q,\omega)}{\omega}  \underset{q\to 0}{=} \frac{ N \kappa}{2}
\label{eq:chi_q}
\end{equation}
with $\kappa$ the compressibility of the system (hereafter we set $\hbar=m=1$, with $m$ the atomic mass), while $S(q,\omega)$ is the dynamic structure factor. 
From the analysis of the measurable signal $F(t)$ of Eq.~(\ref{eq:df}) one can then determine the sound velocities $c_+$ and $c_-$, and the relative contribution
\begin{equation}
 R\equiv \frac{\chi_-}{\chi}= \frac{c^2_+-c^2_\kappa}{c_+^2-c^2_-}
  \label{R}
  \end{equation}
of the lowest (lower sound) mode  to the compressibility sum rule, where we have defined $c_\kappa=\sqrt{\kappa^{-1}}$.  
Analogously, the  contribution of lower sound to the  $f$ sum rule $m_1=\int_0^\infty d\omega\, S(q,\omega)\omega=Nq^2/2$ is given by
\begin{equation}
\frac{m_1^{(-)}}{m_1}= \frac{ c^2_-}{c^2_\kappa} \frac{c^2_+ - c^2_\kappa}{c_+^2-c^2_-}.
\label{eq:fSumRule}
\end{equation}

However, in the vicinity of the superfluid-supersolid phase transition, within the supersolid region, we find that our perturbation excites additional modes and Eqs.~(\ref{R}-\ref{eq:fSumRule}), based on the two-mode approximation, are less accurate.

\paragraph{Hydrodynamic model for supersolidity.}
Since the two sound modes are the Nambu–Goldstone bosons due to the spontaneous breaking of translational symmetry and the $U(1)$ symmetry related to the conserved particle number, the low-energy dynamics of a supersolid exhibits universal features and can be described by hydrodynamics~\cite{AndreevLifshitz,Saalow,Pomeau94,Son,Hydro-Rica2007}. 
In the following we will use the hydrodynamic approach to supersolidity recently elaborated 
by Hofmann and Zwerger~\cite{Hofmann2021}, inspired by the work of Yoo and Dorsey~\cite{Dorsey2010}.
This formulation, applicable to Galilean invariant systems, is particularly suitable to investigate  the behavior
of longitudinal phonons in the presence of a layer structure. This is reasonably well realized  in highly elongated configurations of a dipolar supersolid, where the  droplets effectively play the role of the layers.  Neglecting the effects 
of the strain density coupling included in the general formulation of supersolid hydrodynamics~\cite{josserand2007,Dorsey2010}, the approach,
in this minimal hydrodynamic formulation, provides the following expression for the two sound velocities~\cite{Hofmann2021}
\begin{equation}
c^2_{\pm} = \frac{c^2_\kappa}{2}\left[1+\beta \kappa \pm \sqrt{(1+\beta\kappa)^2-4f_S\beta\kappa}\right],
\label{eq:FirstSecondSound}
\end{equation}
which depends on three fundamental parameters:
the  velocity $c_\kappa$, fixed by  the compressibility parameter $\kappa$; the renormalized layer compressibility modulus $\beta=B/\rho_n$, given by the layer compressibility modulus $B$~\cite{chaikin_lubensky_1995} divided by the normal density $\rho_n=\bar{\rho}-\rho_s$; and the superfluid fraction $f_S=\rho_s/\bar{\rho}$, with $\bar{\rho}$ the average 1D density. In our one-dimensional   structure, $B$ is the only  elastic constant, corresponding to the energy cost due to the change in the separation between the peaks. The relevant parameters $\beta\kappa$ and $f_S$ can be expressed in terms   
 of  the  upper and lower sound velocities according to the relations $ \beta\kappa= (c^2_+ +c^2_-)/c^2_\kappa-1 $
and $f_S\beta\kappa= c^2_+c^2_-/c^4_\kappa$, which directly follow from Eq.~(\ref{eq:FirstSecondSound}).

 Let us now  discuss the consequences of the  hydrodynamic model in different phases of dipolar Bose gases:  

(i) Superfluid phase ($f_S=1$ and $\beta=0$). Only the upper solution (upper sound) of Eq.~(\ref{eq:FirstSecondSound}) is relevant and $c_+=c_\kappa $.

(ii) Supersolid phase ($0<f_s <1$ and $\beta\ne 0$). In this  most interesting case the deviations of the sound speeds from  $c_\kappa$ are    determined   by the dimensionless combination $\beta \kappa$ and by the superfluid fraction $f_S$. In particular, near the transition to the crystal phase, where the superfluid fraction is expected to vanish,  the sound velocities approach the values
  \begin{equation}
   c_+ \to  \sqrt{1+\beta\kappa-\frac{f_S \beta\kappa}{1+\beta\kappa}}c_\kappa , \quad
   c_- \to \sqrt{f_S \frac{\beta \kappa}{1+\beta\kappa}}c_\kappa
   \label{eq:c2limit}
  \end{equation} 
  while the ratio $R$ given by Eq.~(\ref{R}) approaches the value $\beta\kappa/(1+\beta\kappa)$. It is worth noticing the close analogy between Eq.~(\ref{eq:c2limit}) and the dependence of the second sound velocity on the superfluid density predicted by Landau's two-fluid hydrodynamic theory at finite temperature~\cite{landau41} (see also~\cite{BecBook2016}).

  
  It is also interesting to note that  when the combination $\beta\kappa $ becomes very large the lower sound velocity    takes the form $c_- = \sqrt{f_S} c_\kappa$ in the whole supersolid phase, it exhausts the compressibility sum rule, while its relative contribution to the $f$-sum rule [see Eq.~(\ref{eq:fSumRule})] exactly coincides with the superfluid fraction $f_S$.    These results are consistent with the behavior of a superfluid in the presence of an optical lattice, where translational invariance is not broken spontaneously and the upper mode $\omega_+=c_+q$ is replaced by a  gapped excitation.
  
(iii) Crystal phase ($f_s=0$ and $\beta\ne 0$). Only the upper solution survives in this case and the  sound velocity takes the simple expression $c_+=c_\kappa\sqrt{1+\beta\kappa}$. Notice however  that, differently from what happens in the superfluid phase, the upper sound mode, while exhausting the  $f$-sum rule,  does not exhaust the compressibility sum rule, revealing the occurrence of a diffusive mode at zero frequency. Such a mode represents the natural continuation of the lower sound mode beyond the transition to the crystal phase~\cite{Martin1972} and  corresponds to the diffusive permeation mode of a smectic-A
liquid crystal~\cite{Martin1972,Hofmann2021}. The evolution of the lower mode from a propagating to a diffusive one, is analogous to the fate of second sound in a uniform fluid above the superfluid critical temperature (see, e.g., \cite{Vinen70,Hu2010}).

\paragraph{Results.}
The compressibility $\kappa$ of the gas can be extracted from the knowledge of the density changes caused by the static perturbation $-V_0\cos(\varphi)$ according to linear response  theory [see Eq.~(\ref{eq:chi_q})]. 
Another option, which would not require the actual knowledge of $V_0$, is to measure the relative contribution $R$ [see  Eq.~(\ref{R})] of the lower sound mode to the compressibility sum rule through the weights of the beating signal of Eq.~(\ref{eq:df}). 

Our protocol actually measures the static response function $\chi(q)/N$ [see Eqs.~(\ref{eq:df}),(\ref{eq:chi_q})], which coincides with the compressibility $\kappa$  only in the long-wavelength limit $q\to 0$. Because of the finite size of the ring, the lowest accessible value is $q=2\pi/L$, and it is consequently important to control the difference between $\chi(q=2\pi/L)/N$ and the compressibility parameter $\kappa=(\bar{\rho}\,\partial \mu/\partial \bar{\rho})^{-1}$, where $\mu$ is the chemical potential.
In our case the difference turns out to be about $15\%$ in the superfluid phase and up to $30\%$ close to the crystal phase.
For consistency, we have used the values of $\kappa$ given by the ``measured" values $\chi(q=2\pi/L)$. On the other hand, we verified  that  the superfluid fraction $f_S$, obtained by applying the hydrodynamic model to the results of the extended Gross-Pitaevskii simulation employing our protocol,  is much less sensitive to finite size effects. 

\begin{figure}[!ht]
    \includegraphics[width=0.99\linewidth]{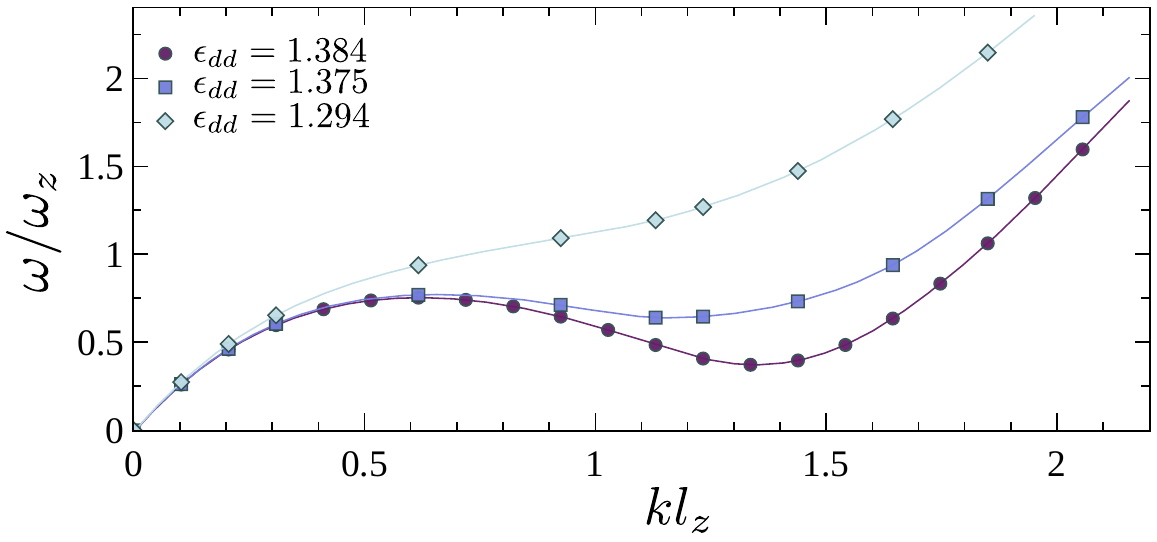}
    \caption{Dispersion relation obtained using the proposed protocol, for three values of $\epsilon_{dd}$ approaching the superfluid-supersolid phase transition. The roton minimum softens near ${k = \sqrt{2}/l_z}$, where ${l_z = \sqrt{\hbar/m\omega_z}}$ is the harmonic oscillator length along the confined direction~\cite{SantosRoton}.}
    \label{fig:1}
\end{figure}
To illustrate the potential of the proposed protocol, in Fig.~\ref{fig:1} we report  the dispersion for larger values of $q$, within the superfluid phase, obtained   by applying  a perturbation proportional to $\cos(n\varphi)$ with $n=1,2, ...$,  giving access to the phonon-maxon-roton dispersion, for which experimental evidence was reported in a superfluid dipolar gas using Bragg spectroscopy~\cite{petter-roton2019}. The figure clearly shows that the  roton minimum becomes more pronounced as one approaches the transition to the supersolid phase, which in our configuration occurs for $\epsilon_{dd} = 1.387$.

\begin{figure}[!ht]
    \includegraphics[width=0.99\linewidth]{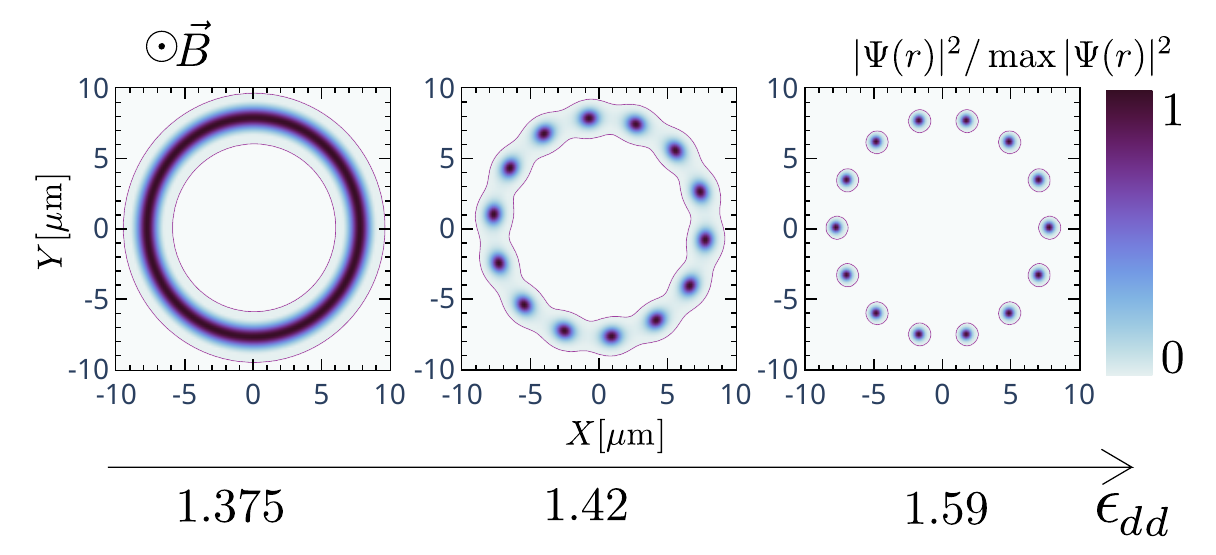}
    \caption{Density plots of $N=80000$ $^{164}$Dy atoms in the superfluid, supersolid and crystal phase (left, middle, and right panel respectively), integrated over the $z$-axis, along which the magnetic field $\vec{B}$ is aligned. Red contours mark $1\%$ of the relative density $|\Psi(\boldsymbol{r})|^2 / \max |\Psi(\boldsymbol{r})|^2$.}
    \label{fig:2}
\end{figure}
In Fig.~\ref{fig:2} we show the density profiles in the ring geometry calculated in the superfluid, supersolid, and crystal phases~\footnote{Ring configurations of similar shape, hosting a supersolid dipolar gas with persistent currents, were considered in \cite{tengstrand2021}.}.
In our simulations, based on the extended Gross-Pitaevskii equation, 
we have considered configurations with the same number of droplets (equal to 14) in both the supersolid and crystal phases.  Actually, exact energy minimization would predict a decrease of the number of droplets  when one approaches the transition to the crystal phase, leading to the small discontinuities  in the resulting values of the observed quantities, which do not however affect the main conclusions of our work. The pinning of the number (and position) of droplets can be achieved by introducing a small additional periodic potential during the initial stage of the supersolid state preparation.
During the time evolution, once the periodic potential is removed, we observe that the number of droplets remains constant. 

\begin{figure}[b]
\includegraphics[width=0.99\linewidth]{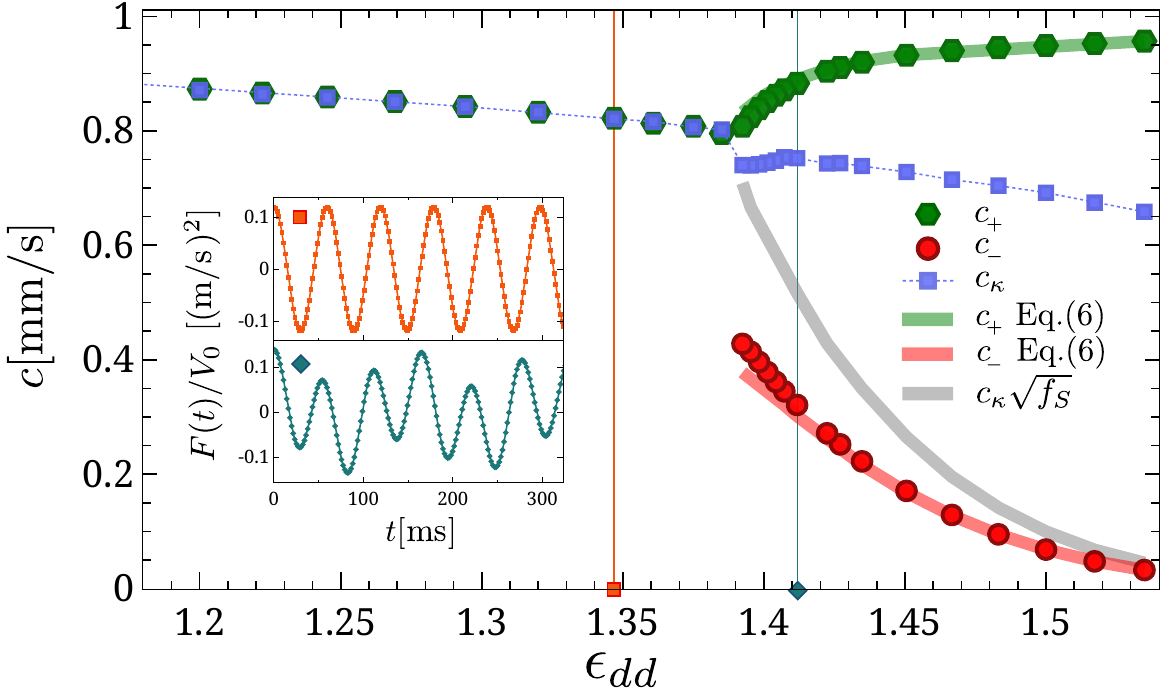}
    \caption{ Sound velocities $c_+$ and $c_-$ determined with the protocol (green hexagons and red circles respectively) across the superfluid-supersolid phase transition. The blue squares correspond to $c_{\kappa}\left(q=2\pi/L\right)$. Both sound speeds are well captured by Eq.~(\ref{eq:c2limit}) (green and red solid lines) even far from the crystal phase. For completeness we also report the values $\sqrt{f_S}c_{\kappa}$ (gray solid line) found for an incompressible lattice (see text). The inset presents the time evolution of $ F$ for two values of $\epsilon_{dd}$ marked with solid lines of corresponding colors. The data points are fitted with one (two) cosines in the superfluid (supersolid) phase with excellent quality. } 
    \label{fig:4} 
\end{figure}
\begin{figure}[h]
\includegraphics[width=0.99\linewidth]{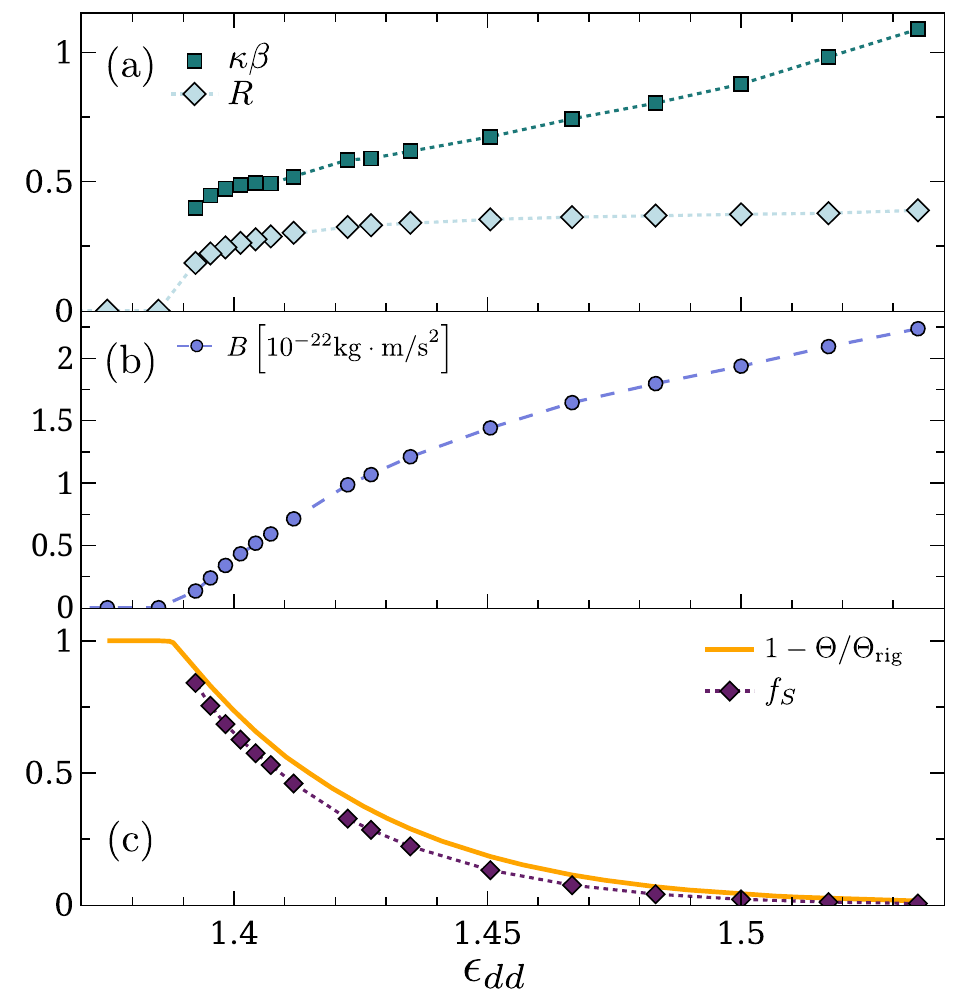}
    \caption{Panel (a) displays dimensionless parameters $\kappa\beta$ calculated using the hydrodynamic model (green squares), and the ratio $R=\chi_-/\chi$ (light blue diamonds) extracted from the time evolution of Eq.~(\ref{eq:df}).  In panel (b) we show the emergence of finite layer compressibility modulus $B$. In panel (c) we compare the extracted value of superfluid fraction $f_S$ (purple diamonds) using the hydrodynamic relations Eq.~(\ref{eq:FirstSecondSound}), with the value determined via the nonclassical fraction of moment of inertia
    Eq.~(\ref{eq:fs}) (orange line).}
    \label{fig:5}
\end{figure}

Figures \ref{fig:4} and \ref{fig:5} report the main results of our work, based on the combined application of the protocol and of the hydrodynamic model of supersolids. In Fig.~\ref{fig:4} we show  the  calculated upper and lower sound velocities as a function of $\epsilon_{dd}$, together with the value of $c_\kappa$, which coincides with the sound velocity in the superfluid phase. The figure clearly reveals the decrease of the lower sound velocity as one approaches the transition to the crystal phase. 
Similar features have been reported in \cite{Blakie2023sounds} obtained for an infinite tube, confirming that our mesoscopic system correctly approaches the thermodynamic limit. We include also the lowest order expressions for the two sound speeds when $f_s\rightarrow 0$, Eq.~(\ref{eq:c2limit}) (red and green continuous lines), which are seen to be in good agreement with the calculated values also when the superfluid density is not that small.
For comparison we also show  the prediction $c_-=\sqrt{f_S}c_\kappa$ (grey continuous line), which would hold in the presence of an optical lattice ($B\to \infty$) and which badly reproduces the actual  values of $c_-$ in the supersolid phase. The standard  hydrodynamic expression $c=\sqrt{f_S}c_\kappa$ was actually successfully employed by Tao \textit{et al.} \cite{Tao2023} and  Chauvaeau \textit{et al.} \cite{Chauveau2023} to extract the value of $f_S$ of a BEC gas in the presence of an optical lattice (not a supersolid).

In Fig.~4(a) we report the  results for  the relevant parameter $\beta\kappa$ of the hydrodynamic model and  the ratio $R$. An interesting outcome of our analysis is that while the speeds of sound, the compressibility and the dimensionless parameter $\beta\kappa$ show a jump at the superfluid-supersolid transition, the contribution of lower sound to the compressibility sum rule, $R$, goes smoothly to zero. The same continuous vanishing is observed for the layer compressibility modulus $B$, as shown in Fig.~4(b). 
In Fig.~4(c) we report the value of the superfluid fraction $f_S$, predicted by the analysis of the sound velocities: $f_S= c^2_+c^2_-/c^2_\kappa(c^2_++c^2_--c^2_\kappa)$. As already pointed out above, we have verified that this value of $f_S$ is very weakly sensitive to finite size effects. Furthermore, repeating the simulations with a larger value of  the phonon wave vector, $q=4\pi/L$, rather than $2\pi/L$, we find that $f_S$ is modified only by a few percent. Since the oscillation frequency has also almost doubled, this would allow working on shorter timescales, thereby reducing the consequences of the finite lifetime of the system.

\paragraph{Moment of inertia and superfluid fraction.}

Both the layer compressibility and the superfluid fraction can, in principle, be calculated from Gross-Pitaevskii theory (see, for example, Josserand et al.~\cite{josserand2007}). 
In fact, the superfluid fraction has been the object of several calculations based on extended Gross-Pitaevskii theory (see, e.g., \cite{Roccuzzo1,Smith2023,Blakie2023sounds}) for dipolar gases showing one-dimensional periodic configurations. In our narrow ring configuration, the superfluid fraction essentially coincides with the non-classical fraction of the moment of inertia
\begin{equation}
f_S\simeq 1-\frac{\Theta}{\Theta_\mathrm{rig}} \; ,
\label{eq:fs}
\end{equation}
which is  reported in Fig.~4(c) (orange line) for the comparison with the value 
 extracted from the calculated sound velocities.
The moment of inertia $\Theta$ is fixed by the value $\langle J_z\rangle$ of   the angular momentum induced by a rotational constraint  of the form $-\Omega J_z$, according to the relation $\Theta=\lim_{\Omega\rightarrow 0} \langle J_z\rangle/\Omega $,  while  $\Theta_\mathrm{rig}=N\langle x^2+y^2\rangle$ is the classical rigid value~\footnote{The energy dependence on angular momentum for such rotating configurations at larger $\Omega$ has recently been explored in \cite{tengstrand2021}. }.
 We also verified that $f_S$ from Eq.~(\ref{eq:fs}) practically coincides with the  rigorous Leggett's upper bound $ 2\pi\bigl[\bar{\rho}\int_0^{2\pi} d\varphi/\rho(\varphi)\bigr]^{-1} \ge f_S$, where 
$\rho(\varphi)$ is the transverse integrated density along the ring~\footnote{The good agreement between Leggett's upper bound and the superfluid fraction calculated in an infinite tube by means of effective mass was recently discussed in \cite{Smith2023}. }. 

The good agreement  between the two predictions shown in Fig.~4(c) reveals the consistency of the extended Gross-Pitaevskii theory  with the hydrodynamic model of  supersolidity.

  
In conclusion, we have suggested (i) a protocol to determine the Goldstone modes of a supersolid dipolar gas confined in a ring and (ii) a way to  identify the relevant parameters of the hydrodynamic theory of supersolids. 

Our work, in particular, paves the way for an experimental determination of the layer compressibility modulus and of the superfluid fraction, based on the measurement of sound velocities. On the theory side, our findings can be used to extract the supersolid hydrodynamic parameters in the thermodynamic limit by using the available estimations of the sound speeds   
in infinite systems \cite{Roccuzzo1,Blakie2023sounds}.



Finally, it is worth noticing that our protocol for the extraction of the speed of sound and of the hydrodynamic parameters is not limited to the dipolar supersolid platform we discuss in this Letter. It can be equally applied to dipolar mixtures, which are expected to show longer lifetimes than single-component dipolar gases~\cite{Ueda2009,Bland2022-dipmix,Li2022-dipmix,scheiermann2023}, and whose spectrum has been very recently addressed in~\cite{kirkby2023excitations}.
Another very interesting system for the application of our protocol are semiconductor dipolar excitons, where ring configurations naturally occur (see, e.g., \cite{Rapaport_2007,Combescot_rev_2017} and references therein), and cluster crystallisation has been observed \cite{ButovNature2002, Andreev2017}.

\paragraph{Acknowledgements.} 
We are grateful to Wilhelm Zwerger for many illuminating comments concerning the hydrodynamic model for supersolids when applied to configurations with  density modulation along a single
direction. Stimulating discussions with Giovanni Modugno and William Phillips  are also acknowledged. We acknowledge support from the Provincia Autonoma di Trento, the Q@TN (the
joint lab between University of Trento, FBK-Fondazione
Bruno Kessler, INFN-National Institute for Nuclear
Physics and CNR-National Research Council), the Italian MIUR under the PRIN2017 project
CEnTraL (Protocol Number 20172H2SC4), and from the CINECA through the award under the ISCRA initiative, for the availability of HPC resources.


\bibliography{bibliosupersolid}

\begin{thebibliography}{52}%
\makeatletter
\providecommand \@ifxundefined [1]{%
 \@ifx{#1\undefined}
}%
\providecommand \@ifnum [1]{%
 \ifnum #1\expandafter \@firstoftwo
 \else \expandafter \@secondoftwo
 \fi
}%
\providecommand \@ifx [1]{%
 \ifx #1\expandafter \@firstoftwo
 \else \expandafter \@secondoftwo
 \fi
}%
\providecommand \natexlab [1]{#1}%
\providecommand \enquote  [1]{``#1''}%
\providecommand \bibnamefont  [1]{#1}%
\providecommand \bibfnamefont [1]{#1}%
\providecommand \citenamefont [1]{#1}%
\providecommand \href@noop [0]{\@secondoftwo}%
\providecommand \href [0]{\begingroup \@sanitize@url \@href}%
\providecommand \@href[1]{\@@startlink{#1}\@@href}%
\providecommand \@@href[1]{\endgroup#1\@@endlink}%
\providecommand \@sanitize@url [0]{\catcode `\\12\catcode `\$12\catcode `\&12\catcode `\#12\catcode `\^12\catcode `\_12\catcode `\%12\relax}%
\providecommand \@@startlink[1]{}%
\providecommand \@@endlink[0]{}%
\providecommand \url  [0]{\begingroup\@sanitize@url \@url }%
\providecommand \@url [1]{\endgroup\@href {#1}{\urlprefix }}%
\providecommand \urlprefix  [0]{URL }%
\providecommand \Eprint [0]{\href }%
\providecommand \doibase [0]{https://doi.org/}%
\providecommand \selectlanguage [0]{\@gobble}%
\providecommand \bibinfo  [0]{\@secondoftwo}%
\providecommand \bibfield  [0]{\@secondoftwo}%
\providecommand \translation [1]{[#1]}%
\providecommand \BibitemOpen [0]{}%
\providecommand \bibitemStop [0]{}%
\providecommand \bibitemNoStop [0]{.\EOS\space}%
\providecommand \EOS [0]{\spacefactor3000\relax}%
\providecommand \BibitemShut  [1]{\csname bibitem#1\endcsname}%
\let\auto@bib@innerbib\@empty
\bibitem [{\citenamefont {Andreev}\ and\ \citenamefont {Lifshitz}(1969)}]{AndreevLifshitz}%
  \BibitemOpen
  \bibfield  {author} {\bibinfo {author} {\bibfnamefont {A.~F.}\ \bibnamefont {Andreev}}\ and\ \bibinfo {author} {\bibfnamefont {I.~M.}\ \bibnamefont {Lifshitz}},\ }\bibfield  {title} {\bibinfo {title} {Quantum theory of defects in crystals},\ }\href {http://www.jetp.ras.ru/cgi-bin/e/index/e/29/6/p1107?a=list} {\bibfield  {journal} {\bibinfo  {journal} {Sov. Phys. JETP}\ }\textbf {\bibinfo {volume} {29}},\ \bibinfo {pages} {1107} (\bibinfo {year} {1969})}\BibitemShut {NoStop}%
\bibitem [{\citenamefont {Saslow}(1977)}]{Saalow}%
  \BibitemOpen
  \bibfield  {author} {\bibinfo {author} {\bibfnamefont {W.~M.}\ \bibnamefont {Saslow}},\ }\bibfield  {title} {\bibinfo {title} {Microscopic and hydrodynamic theory of superfluidity in periodic solids},\ }\href {https://doi.org/10.1103/PhysRevB.15.173} {\bibfield  {journal} {\bibinfo  {journal} {Phys. Rev. B}\ }\textbf {\bibinfo {volume} {15}},\ \bibinfo {pages} {173} (\bibinfo {year} {1977})}\BibitemShut {NoStop}%
\bibitem [{\citenamefont {Son}(2005)}]{Son}%
  \BibitemOpen
  \bibfield  {author} {\bibinfo {author} {\bibfnamefont {D.~T.}\ \bibnamefont {Son}},\ }\bibfield  {title} {\bibinfo {title} {Effective lagrangian and topological interactions in supersolids},\ }\href {https://doi.org/10.1103/PhysRevLett.94.175301} {\bibfield  {journal} {\bibinfo  {journal} {Phys. Rev. Lett.}\ }\textbf {\bibinfo {volume} {94}},\ \bibinfo {pages} {175301} (\bibinfo {year} {2005})}\BibitemShut {NoStop}%
\bibitem [{\citenamefont {Josserand}\ \emph {et~al.}(2007{\natexlab{a}})\citenamefont {Josserand}, \citenamefont {Pomeau},\ and\ \citenamefont {Rica}}]{Hydro-Rica2007}%
  \BibitemOpen
  \bibfield  {author} {\bibinfo {author} {\bibfnamefont {C.}~\bibnamefont {Josserand}}, \bibinfo {author} {\bibfnamefont {Y.}~\bibnamefont {Pomeau}},\ and\ \bibinfo {author} {\bibfnamefont {S.}~\bibnamefont {Rica}},\ }\bibfield  {title} {\bibinfo {title} {Coexistence of ordinary elasticity and superfluidity in a model of a defect-free supersolid},\ }\href {https://doi.org/10.1103/PhysRevLett.98.195301} {\bibfield  {journal} {\bibinfo  {journal} {Phys. Rev. Lett.}\ }\textbf {\bibinfo {volume} {98}},\ \bibinfo {pages} {195301} (\bibinfo {year} {2007}{\natexlab{a}})}\BibitemShut {NoStop}%
\bibitem [{\citenamefont {Yoo}\ and\ \citenamefont {Dorsey}(2010)}]{Dorsey2010}%
  \BibitemOpen
  \bibfield  {author} {\bibinfo {author} {\bibfnamefont {C.-D.}\ \bibnamefont {Yoo}}\ and\ \bibinfo {author} {\bibfnamefont {A.~T.}\ \bibnamefont {Dorsey}},\ }\bibfield  {title} {\bibinfo {title} {Hydrodynamic theory of supersolids: Variational principle, effective lagrangian, and density-density correlation function},\ }\href {https://doi.org/10.1103/PhysRevB.81.134518} {\bibfield  {journal} {\bibinfo  {journal} {Phys. Rev. B}\ }\textbf {\bibinfo {volume} {81}},\ \bibinfo {pages} {134518} (\bibinfo {year} {2010})}\BibitemShut {NoStop}%
\bibitem [{\citenamefont {Hofmann}\ and\ \citenamefont {Zwerger}(2021)}]{Hofmann2021}%
  \BibitemOpen
  \bibfield  {author} {\bibinfo {author} {\bibfnamefont {J.}~\bibnamefont {Hofmann}}\ and\ \bibinfo {author} {\bibfnamefont {W.}~\bibnamefont {Zwerger}},\ }\bibfield  {title} {\bibinfo {title} {Hydrodynamics of a superfluid smectic},\ }\href {https://doi.org/10.1088/1742-5468/abe598} {\bibfield  {journal} {\bibinfo  {journal} {Journal of Statistical Mechanics: Theory and Experiment}\ }\textbf {\bibinfo {volume} {2021}},\ \bibinfo {pages} {033104} (\bibinfo {year} {2021})}\BibitemShut {NoStop}%
\bibitem [{\citenamefont {Saccani}\ \emph {et~al.}(2012)\citenamefont {Saccani}, \citenamefont {Moroni},\ and\ \citenamefont {Boninsegni}}]{Boninsegni_SS2012}%
  \BibitemOpen
  \bibfield  {author} {\bibinfo {author} {\bibfnamefont {S.}~\bibnamefont {Saccani}}, \bibinfo {author} {\bibfnamefont {S.}~\bibnamefont {Moroni}},\ and\ \bibinfo {author} {\bibfnamefont {M.}~\bibnamefont {Boninsegni}},\ }\bibfield  {title} {\bibinfo {title} {Excitation spectrum of a supersolid},\ }\href {https://doi.org/10.1103/PhysRevLett.108.175301} {\bibfield  {journal} {\bibinfo  {journal} {Phys. Rev. Lett.}\ }\textbf {\bibinfo {volume} {108}},\ \bibinfo {pages} {175301} (\bibinfo {year} {2012})}\BibitemShut {NoStop}%
\bibitem [{\citenamefont {Macr\`{\i}}\ \emph {et~al.}(2013)\citenamefont {Macr\`{\i}}, \citenamefont {Maucher}, \citenamefont {Cinti},\ and\ \citenamefont {Pohl}}]{Macri_SS}%
  \BibitemOpen
  \bibfield  {author} {\bibinfo {author} {\bibfnamefont {T.}~\bibnamefont {Macr\`{\i}}}, \bibinfo {author} {\bibfnamefont {F.}~\bibnamefont {Maucher}}, \bibinfo {author} {\bibfnamefont {F.}~\bibnamefont {Cinti}},\ and\ \bibinfo {author} {\bibfnamefont {T.}~\bibnamefont {Pohl}},\ }\bibfield  {title} {\bibinfo {title} {Elementary excitations of ultracold soft-core bosons across the superfluid-supersolid phase transition},\ }\href {https://doi.org/10.1103/PhysRevA.87.061602} {\bibfield  {journal} {\bibinfo  {journal} {Phys. Rev. A}\ }\textbf {\bibinfo {volume} {87}},\ \bibinfo {pages} {061602} (\bibinfo {year} {2013})}\BibitemShut {NoStop}%
\bibitem [{\citenamefont {Ancilotto}\ \emph {et~al.}(2013)\citenamefont {Ancilotto}, \citenamefont {Rossi},\ and\ \citenamefont {Toigo}}]{Ancilotto2013}%
  \BibitemOpen
  \bibfield  {author} {\bibinfo {author} {\bibfnamefont {F.}~\bibnamefont {Ancilotto}}, \bibinfo {author} {\bibfnamefont {M.}~\bibnamefont {Rossi}},\ and\ \bibinfo {author} {\bibfnamefont {F.}~\bibnamefont {Toigo}},\ }\bibfield  {title} {\bibinfo {title} {Supersolid structure and excitation spectrum of soft-core bosons in three dimensions},\ }\href {https://doi.org/10.1103/PhysRevA.88.033618} {\bibfield  {journal} {\bibinfo  {journal} {Phys. Rev. A}\ }\textbf {\bibinfo {volume} {88}},\ \bibinfo {pages} {033618} (\bibinfo {year} {2013})}\BibitemShut {NoStop}%
\bibitem [{\citenamefont {Recati}\ and\ \citenamefont {Stringari}(2022)}]{ARCMP}%
  \BibitemOpen
  \bibfield  {author} {\bibinfo {author} {\bibfnamefont {A.}~\bibnamefont {Recati}}\ and\ \bibinfo {author} {\bibfnamefont {S.}~\bibnamefont {Stringari}},\ }\bibfield  {title} {\bibinfo {title} {Coherently coupled mixtures of ultracold atomic gases},\ }\href {https://doi.org/10.1146/annurev-conmatphys-031820-121316} {\bibfield  {journal} {\bibinfo  {journal} {Annual Review of Condensed Matter Physics}\ }\textbf {\bibinfo {volume} {13}},\ \bibinfo {pages} {407} (\bibinfo {year} {2022})},\ \Eprint {https://arxiv.org/abs/https://doi.org/10.1146/annurev-conmatphys-031820-121316} {https://doi.org/10.1146/annurev-conmatphys-031820-121316} \BibitemShut {NoStop}%
\bibitem [{\citenamefont {Martone}(2023)}]{MartoneReview}%
  \BibitemOpen
  \bibfield  {author} {\bibinfo {author} {\bibfnamefont {G.~I.}\ \bibnamefont {Martone}},\ }\bibfield  {title} {\bibinfo {title} {Bose-einstein condensates with raman-induced spin-orbit coupling: An overview},\ }\href {https://doi.org/10.1209/0295-5075/ace2e8} {\bibfield  {journal} {\bibinfo  {journal} {Europhysics Letters}\ }\textbf {\bibinfo {volume} {143}},\ \bibinfo {pages} {25001} (\bibinfo {year} {2023})}\BibitemShut {NoStop}%
\bibitem [{\citenamefont {Masalaeva}\ \emph {et~al.}(2023)\citenamefont {Masalaeva}, \citenamefont {Ritsch},\ and\ \citenamefont {Mivehvar}}]{natalia2023}%
  \BibitemOpen
  \bibfield  {author} {\bibinfo {author} {\bibfnamefont {N.}~\bibnamefont {Masalaeva}}, \bibinfo {author} {\bibfnamefont {H.}~\bibnamefont {Ritsch}},\ and\ \bibinfo {author} {\bibfnamefont {F.}~\bibnamefont {Mivehvar}},\ }\bibfield  {title} {\bibinfo {title} {Tuning photon-mediated interactions in a multimode cavity: From supersolid to insulating droplets hosting phononic excitations},\ }\href {https://doi.org/10.1103/PhysRevLett.131.173401} {\bibfield  {journal} {\bibinfo  {journal} {Phys. Rev. Lett.}\ }\textbf {\bibinfo {volume} {131}},\ \bibinfo {pages} {173401} (\bibinfo {year} {2023})}\BibitemShut {NoStop}%
\bibitem [{\citenamefont {Roccuzzo}\ and\ \citenamefont {Ancilotto}(2019)}]{Roccuzzo1}%
  \BibitemOpen
  \bibfield  {author} {\bibinfo {author} {\bibfnamefont {S.~M.}\ \bibnamefont {Roccuzzo}}\ and\ \bibinfo {author} {\bibfnamefont {F.}~\bibnamefont {Ancilotto}},\ }\bibfield  {title} {\bibinfo {title} {Supersolid behavior of a dipolar bose-einstein condensate confined in a tube},\ }\href {https://doi.org/10.1103/PhysRevA.99.041601} {\bibfield  {journal} {\bibinfo  {journal} {Phys. Rev. A}\ }\textbf {\bibinfo {volume} {99}},\ \bibinfo {pages} {041601} (\bibinfo {year} {2019})}\BibitemShut {NoStop}%
\bibitem [{\citenamefont {Recati}\ and\ \citenamefont {Stringari}(2023)}]{NaturePersp}%
  \BibitemOpen
  \bibfield  {author} {\bibinfo {author} {\bibfnamefont {A.}~\bibnamefont {Recati}}\ and\ \bibinfo {author} {\bibfnamefont {S.}~\bibnamefont {Stringari}},\ }\bibfield  {title} {\bibinfo {title} {Supersolidity in ultracold dipolar gases},\ }\href {https://doi.org/10.1038/s42254-023-00648-2} {\bibfield  {journal} {\bibinfo  {journal} {Nature Reviews Physics}\ }\textbf {\bibinfo {volume} {5}},\ \bibinfo {pages} {735} (\bibinfo {year} {2023})}\BibitemShut {NoStop}%
\bibitem [{\citenamefont {Tanzi}\ \emph {et~al.}(2019)\citenamefont {Tanzi}, \citenamefont {Roccuzzo}, \citenamefont {Lucioni}, \citenamefont {Fam{\`{a}}}, \citenamefont {Fioretti}, \citenamefont {Gabbanini}, \citenamefont {Modugno}, \citenamefont {Recati},\ and\ \citenamefont {Stringari}}]{F2}%
  \BibitemOpen
  \bibfield  {author} {\bibinfo {author} {\bibfnamefont {L.}~\bibnamefont {Tanzi}}, \bibinfo {author} {\bibfnamefont {S.~M.}\ \bibnamefont {Roccuzzo}}, \bibinfo {author} {\bibfnamefont {E.}~\bibnamefont {Lucioni}}, \bibinfo {author} {\bibfnamefont {F.}~\bibnamefont {Fam{\`{a}}}}, \bibinfo {author} {\bibfnamefont {A.}~\bibnamefont {Fioretti}}, \bibinfo {author} {\bibfnamefont {C.}~\bibnamefont {Gabbanini}}, \bibinfo {author} {\bibfnamefont {G.}~\bibnamefont {Modugno}}, \bibinfo {author} {\bibfnamefont {A.}~\bibnamefont {Recati}},\ and\ \bibinfo {author} {\bibfnamefont {S.}~\bibnamefont {Stringari}},\ }\bibfield  {title} {\bibinfo {title} {{Supersolid symmetry breaking from compressional oscillations in a dipolar quantum gas}},\ }\href {https://doi.org/10.1038/s41586-019-1568-6} {\bibfield  {journal} {\bibinfo  {journal} {Nature}\ }\textbf {\bibinfo {volume} {574}},\ \bibinfo {pages} {382} (\bibinfo {year} {2019})}\BibitemShut {NoStop}%
\bibitem [{\citenamefont {Guo}\ \emph {et~al.}(2019)\citenamefont {Guo}, \citenamefont {B{\"{o}}ttcher}, \citenamefont {Hertkorn}, \citenamefont {Schmidt}, \citenamefont {Wenzel}, \citenamefont {B{\"{u}}chler}, \citenamefont {Langen},\ and\ \citenamefont {Pfau}}]{S2}%
  \BibitemOpen
  \bibfield  {author} {\bibinfo {author} {\bibfnamefont {M.}~\bibnamefont {Guo}}, \bibinfo {author} {\bibfnamefont {F.}~\bibnamefont {B{\"{o}}ttcher}}, \bibinfo {author} {\bibfnamefont {J.}~\bibnamefont {Hertkorn}}, \bibinfo {author} {\bibfnamefont {J.-N.}\ \bibnamefont {Schmidt}}, \bibinfo {author} {\bibfnamefont {M.}~\bibnamefont {Wenzel}}, \bibinfo {author} {\bibfnamefont {H.~P.}\ \bibnamefont {B{\"{u}}chler}}, \bibinfo {author} {\bibfnamefont {T.}~\bibnamefont {Langen}},\ and\ \bibinfo {author} {\bibfnamefont {T.}~\bibnamefont {Pfau}},\ }\bibfield  {title} {\bibinfo {title} {{The low-energy Goldstone mode in a trapped dipolar supersolid}},\ }\href {https://doi.org/10.1038/s41586-019-1569-5} {\bibfield  {journal} {\bibinfo  {journal} {Nature}\ }\textbf {\bibinfo {volume} {574}},\ \bibinfo {pages} {386} (\bibinfo {year} {2019})}\BibitemShut {NoStop}%
\bibitem [{\citenamefont {Natale}\ \emph {et~al.}(2019)\citenamefont {Natale}, \citenamefont {van Bijnen}, \citenamefont {Patscheider}, \citenamefont {Petter}, \citenamefont {Mark}, \citenamefont {Chomaz},\ and\ \citenamefont {Ferlaino}}]{I2}%
  \BibitemOpen
  \bibfield  {author} {\bibinfo {author} {\bibfnamefont {G.}~\bibnamefont {Natale}}, \bibinfo {author} {\bibfnamefont {R.~M.~W.}\ \bibnamefont {van Bijnen}}, \bibinfo {author} {\bibfnamefont {A.}~\bibnamefont {Patscheider}}, \bibinfo {author} {\bibfnamefont {D.}~\bibnamefont {Petter}}, \bibinfo {author} {\bibfnamefont {M.~J.}\ \bibnamefont {Mark}}, \bibinfo {author} {\bibfnamefont {L.}~\bibnamefont {Chomaz}},\ and\ \bibinfo {author} {\bibfnamefont {F.}~\bibnamefont {Ferlaino}},\ }\bibfield  {title} {\bibinfo {title} {Excitation spectrum of a trapped dipolar supersolid and its experimental evidence},\ }\href {https://doi.org/10.1103/PhysRevLett.123.050402} {\bibfield  {journal} {\bibinfo  {journal} {Phys. Rev. Lett.}\ }\textbf {\bibinfo {volume} {123}},\ \bibinfo {pages} {050402} (\bibinfo {year} {2019})}\BibitemShut {NoStop}%
\bibitem [{\citenamefont {Blakie}\ \emph {et~al.}(2023)\citenamefont {Blakie}, \citenamefont {Chomaz}, \citenamefont {Baillie},\ and\ \citenamefont {Ferlaino}}]{Blakie2023sounds}%
  \BibitemOpen
  \bibfield  {author} {\bibinfo {author} {\bibfnamefont {P.~B.}\ \bibnamefont {Blakie}}, \bibinfo {author} {\bibfnamefont {L.}~\bibnamefont {Chomaz}}, \bibinfo {author} {\bibfnamefont {D.}~\bibnamefont {Baillie}},\ and\ \bibinfo {author} {\bibfnamefont {F.}~\bibnamefont {Ferlaino}},\ }\bibfield  {title} {\bibinfo {title} {Compressibility and speeds of sound across the superfluid-to-supersolid phase transition of an elongated dipolar gas},\ }\href {https://doi.org/10.1103/PhysRevResearch.5.033161} {\bibfield  {journal} {\bibinfo  {journal} {Phys. Rev. Res.}\ }\textbf {\bibinfo {volume} {5}},\ \bibinfo {pages} {033161} (\bibinfo {year} {2023})}\BibitemShut {NoStop}%
\bibitem [{\citenamefont {Lu}\ \emph {et~al.}(2010)\citenamefont {Lu}, \citenamefont {Lu}, \citenamefont {Zhang}, \citenamefont {Qiu}, \citenamefont {Pu},\ and\ \citenamefont {Yi}}]{Lu-preEdge2010}%
  \BibitemOpen
  \bibfield  {author} {\bibinfo {author} {\bibfnamefont {H.-Y.}\ \bibnamefont {Lu}}, \bibinfo {author} {\bibfnamefont {H.}~\bibnamefont {Lu}}, \bibinfo {author} {\bibfnamefont {J.-N.}\ \bibnamefont {Zhang}}, \bibinfo {author} {\bibfnamefont {R.-Z.}\ \bibnamefont {Qiu}}, \bibinfo {author} {\bibfnamefont {H.}~\bibnamefont {Pu}},\ and\ \bibinfo {author} {\bibfnamefont {S.}~\bibnamefont {Yi}},\ }\bibfield  {title} {\bibinfo {title} {Spatial density oscillations in trapped dipolar condensates},\ }\href {https://doi.org/10.1103/PhysRevA.82.023622} {\bibfield  {journal} {\bibinfo  {journal} {Phys. Rev. A}\ }\textbf {\bibinfo {volume} {82}},\ \bibinfo {pages} {023622} (\bibinfo {year} {2010})}\BibitemShut {NoStop}%
\bibitem [{\citenamefont {Roccuzzo}\ \emph {et~al.}(2022)\citenamefont {Roccuzzo}, \citenamefont {Stringari},\ and\ \citenamefont {Recati}}]{roccuzzoEdge}%
  \BibitemOpen
  \bibfield  {author} {\bibinfo {author} {\bibfnamefont {S.~M.}\ \bibnamefont {Roccuzzo}}, \bibinfo {author} {\bibfnamefont {S.}~\bibnamefont {Stringari}},\ and\ \bibinfo {author} {\bibfnamefont {A.}~\bibnamefont {Recati}},\ }\bibfield  {title} {\bibinfo {title} {Supersolid edge and bulk phases of a dipolar quantum gas in a box},\ }\href {https://doi.org/10.1103/PhysRevResearch.4.013086} {\bibfield  {journal} {\bibinfo  {journal} {Phys. Rev. Res.}\ }\textbf {\bibinfo {volume} {4}},\ \bibinfo {pages} {013086} (\bibinfo {year} {2022})}\BibitemShut {NoStop}%
\bibitem [{\citenamefont {Amico}\ \emph {et~al.}(2021)\citenamefont {Amico}, \citenamefont {Boshier}, \citenamefont {Birkl}, \citenamefont {Minguzzi}, \citenamefont {Miniatura}, \citenamefont {Kwek}, \citenamefont {Aghamalyan}, \citenamefont {Ahufinger}, \citenamefont {Anderson}, \citenamefont {Andrei}, \citenamefont {Arnold}, \citenamefont {Baker}, \citenamefont {Bell}, \citenamefont {Bland}, \citenamefont {Brantut}, \citenamefont {Cassettari}, \citenamefont {Chetcuti}, \citenamefont {Chevy}, \citenamefont {Citro}, \citenamefont {De~Palo}, \citenamefont {Dumke}, \citenamefont {Edwards}, \citenamefont {Folman}, \citenamefont {Fortagh}, \citenamefont {Gardiner}, \citenamefont {Garraway}, \citenamefont {Gauthier}, \citenamefont {Günther}, \citenamefont {Haug}, \citenamefont {Hufnagel}, \citenamefont {Keil}, \citenamefont {Ireland}, \citenamefont {Lebrat}, \citenamefont {Li}, \citenamefont {Longchambon}, \citenamefont {Mompart}, \citenamefont {Morsch}, \citenamefont {Naldesi}, \citenamefont {Neely},
  \citenamefont {Olshanii}, \citenamefont {Orignac}, \citenamefont {Pandey}, \citenamefont {Pérez-Obiol}, \citenamefont {Perrin}, \citenamefont {Piroli}, \citenamefont {Polo}, \citenamefont {Pritchard}, \citenamefont {Proukakis}, \citenamefont {Rylands}, \citenamefont {Rubinsztein-Dunlop}, \citenamefont {Scazza}, \citenamefont {Stringari}, \citenamefont {Tosto}, \citenamefont {Trombettoni}, \citenamefont {Victorin}, \citenamefont {Klitzing}, \citenamefont {Wilkowski}, \citenamefont {Xhani},\ and\ \citenamefont {Yakimenko}}]{RoadmapAtomTronic}%
  \BibitemOpen
  \bibfield  {author} {\bibinfo {author} {\bibfnamefont {L.}~\bibnamefont {Amico}}, \bibinfo {author} {\bibfnamefont {M.}~\bibnamefont {Boshier}}, \bibinfo {author} {\bibfnamefont {G.}~\bibnamefont {Birkl}}, \bibinfo {author} {\bibfnamefont {A.}~\bibnamefont {Minguzzi}}, \bibinfo {author} {\bibfnamefont {C.}~\bibnamefont {Miniatura}}, \bibinfo {author} {\bibfnamefont {L.-C.}\ \bibnamefont {Kwek}}, \bibinfo {author} {\bibfnamefont {D.}~\bibnamefont {Aghamalyan}}, \bibinfo {author} {\bibfnamefont {V.}~\bibnamefont {Ahufinger}}, \bibinfo {author} {\bibfnamefont {D.}~\bibnamefont {Anderson}}, \bibinfo {author} {\bibfnamefont {N.}~\bibnamefont {Andrei}}, \bibinfo {author} {\bibfnamefont {A.~S.}\ \bibnamefont {Arnold}}, \bibinfo {author} {\bibfnamefont {M.}~\bibnamefont {Baker}}, \bibinfo {author} {\bibfnamefont {T.~A.}\ \bibnamefont {Bell}}, \bibinfo {author} {\bibfnamefont {T.}~\bibnamefont {Bland}}, \bibinfo {author} {\bibfnamefont {J.~P.}\ \bibnamefont {Brantut}}, \bibinfo {author} {\bibfnamefont
  {D.}~\bibnamefont {Cassettari}}, \bibinfo {author} {\bibfnamefont {W.~J.}\ \bibnamefont {Chetcuti}}, \bibinfo {author} {\bibfnamefont {F.}~\bibnamefont {Chevy}}, \bibinfo {author} {\bibfnamefont {R.}~\bibnamefont {Citro}}, \bibinfo {author} {\bibfnamefont {S.}~\bibnamefont {De~Palo}}, \bibinfo {author} {\bibfnamefont {R.}~\bibnamefont {Dumke}}, \bibinfo {author} {\bibfnamefont {M.}~\bibnamefont {Edwards}}, \bibinfo {author} {\bibfnamefont {R.}~\bibnamefont {Folman}}, \bibinfo {author} {\bibfnamefont {J.}~\bibnamefont {Fortagh}}, \bibinfo {author} {\bibfnamefont {S.~A.}\ \bibnamefont {Gardiner}}, \bibinfo {author} {\bibfnamefont {B.~M.}\ \bibnamefont {Garraway}}, \bibinfo {author} {\bibfnamefont {G.}~\bibnamefont {Gauthier}}, \bibinfo {author} {\bibfnamefont {A.}~\bibnamefont {Günther}}, \bibinfo {author} {\bibfnamefont {T.}~\bibnamefont {Haug}}, \bibinfo {author} {\bibfnamefont {C.}~\bibnamefont {Hufnagel}}, \bibinfo {author} {\bibfnamefont {M.}~\bibnamefont {Keil}}, \bibinfo {author} {\bibfnamefont
  {P.}~\bibnamefont {Ireland}}, \bibinfo {author} {\bibfnamefont {M.}~\bibnamefont {Lebrat}}, \bibinfo {author} {\bibfnamefont {W.}~\bibnamefont {Li}}, \bibinfo {author} {\bibfnamefont {L.}~\bibnamefont {Longchambon}}, \bibinfo {author} {\bibfnamefont {J.}~\bibnamefont {Mompart}}, \bibinfo {author} {\bibfnamefont {O.}~\bibnamefont {Morsch}}, \bibinfo {author} {\bibfnamefont {P.}~\bibnamefont {Naldesi}}, \bibinfo {author} {\bibfnamefont {T.~W.}\ \bibnamefont {Neely}}, \bibinfo {author} {\bibfnamefont {M.}~\bibnamefont {Olshanii}}, \bibinfo {author} {\bibfnamefont {E.}~\bibnamefont {Orignac}}, \bibinfo {author} {\bibfnamefont {S.}~\bibnamefont {Pandey}}, \bibinfo {author} {\bibfnamefont {A.}~\bibnamefont {Pérez-Obiol}}, \bibinfo {author} {\bibfnamefont {H.}~\bibnamefont {Perrin}}, \bibinfo {author} {\bibfnamefont {L.}~\bibnamefont {Piroli}}, \bibinfo {author} {\bibfnamefont {J.}~\bibnamefont {Polo}}, \bibinfo {author} {\bibfnamefont {A.~L.}\ \bibnamefont {Pritchard}}, \bibinfo {author} {\bibfnamefont {N.~P.}\
  \bibnamefont {Proukakis}}, \bibinfo {author} {\bibfnamefont {C.}~\bibnamefont {Rylands}}, \bibinfo {author} {\bibfnamefont {H.}~\bibnamefont {Rubinsztein-Dunlop}}, \bibinfo {author} {\bibfnamefont {F.}~\bibnamefont {Scazza}}, \bibinfo {author} {\bibfnamefont {S.}~\bibnamefont {Stringari}}, \bibinfo {author} {\bibfnamefont {F.}~\bibnamefont {Tosto}}, \bibinfo {author} {\bibfnamefont {A.}~\bibnamefont {Trombettoni}}, \bibinfo {author} {\bibfnamefont {N.}~\bibnamefont {Victorin}}, \bibinfo {author} {\bibfnamefont {W.~v.}\ \bibnamefont {Klitzing}}, \bibinfo {author} {\bibfnamefont {D.}~\bibnamefont {Wilkowski}}, \bibinfo {author} {\bibfnamefont {K.}~\bibnamefont {Xhani}},\ and\ \bibinfo {author} {\bibfnamefont {A.}~\bibnamefont {Yakimenko}},\ }\bibfield  {title} {\bibinfo {title} {{Roadmap on Atomtronics: State of the art and perspective}},\ }\href {https://doi.org/10.1116/5.0026178} {\bibfield  {journal} {\bibinfo  {journal} {AVS Quantum Science}\ }\textbf {\bibinfo {volume} {3}},\ \bibinfo {pages} {039201}
  (\bibinfo {year} {2021})}\BibitemShut {NoStop}%
\bibitem [{\citenamefont {Amico}\ \emph {et~al.}(2022)\citenamefont {Amico}, \citenamefont {Anderson}, \citenamefont {Boshier}, \citenamefont {Brantut}, \citenamefont {Kwek}, \citenamefont {Minguzzi},\ and\ \citenamefont {von Klitzing}}]{Amico22}%
  \BibitemOpen
  \bibfield  {author} {\bibinfo {author} {\bibfnamefont {L.}~\bibnamefont {Amico}}, \bibinfo {author} {\bibfnamefont {D.}~\bibnamefont {Anderson}}, \bibinfo {author} {\bibfnamefont {M.}~\bibnamefont {Boshier}}, \bibinfo {author} {\bibfnamefont {J.-P.}\ \bibnamefont {Brantut}}, \bibinfo {author} {\bibfnamefont {L.-C.}\ \bibnamefont {Kwek}}, \bibinfo {author} {\bibfnamefont {A.}~\bibnamefont {Minguzzi}},\ and\ \bibinfo {author} {\bibfnamefont {W.}~\bibnamefont {von Klitzing}},\ }\bibfield  {title} {\bibinfo {title} {Colloquium: Atomtronic circuits: From many-body physics to quantum technologies},\ }\href {https://doi.org/10.1103/RevModPhys.94.041001} {\bibfield  {journal} {\bibinfo  {journal} {Rev. Mod. Phys.}\ }\textbf {\bibinfo {volume} {94}},\ \bibinfo {pages} {041001} (\bibinfo {year} {2022})}\BibitemShut {NoStop}%
\bibitem [{\citenamefont {Leggett}(1998)}]{Leggett1998}%
  \BibitemOpen
  \bibfield  {author} {\bibinfo {author} {\bibfnamefont {A.~J.}\ \bibnamefont {Leggett}},\ }\bibfield  {title} {\bibinfo {title} {On the superfluid fraction of an arbitrary many-body system at {$T=0$}},\ }\href {https://doi.org/10.1023/B:JOSS.0000033170.38619.6c} {\bibfield  {journal} {\bibinfo  {journal} {J. Stat. Phys.}\ }\textbf {\bibinfo {volume} {93}},\ \bibinfo {pages} {927} (\bibinfo {year} {1998})}\BibitemShut {NoStop}%
\bibitem [{Note1()}]{Note1}%
  \BibitemOpen
  \bibinfo {note} {The cloud's shape is influenced by long-ranged dipolar repulsion effectively increasing its radius.}\BibitemShut {Stop}%
\bibitem [{\citenamefont {W\"achtler}\ and\ \citenamefont {Santos}(2016)}]{Wachter2016}%
  \BibitemOpen
  \bibfield  {author} {\bibinfo {author} {\bibfnamefont {F.}~\bibnamefont {W\"achtler}}\ and\ \bibinfo {author} {\bibfnamefont {L.}~\bibnamefont {Santos}},\ }\bibfield  {title} {\bibinfo {title} {Quantum filaments in dipolar bose-einstein condensates},\ }\href {https://doi.org/10.1103/PhysRevA.93.061603} {\bibfield  {journal} {\bibinfo  {journal} {Phys. Rev. A}\ }\textbf {\bibinfo {volume} {93}},\ \bibinfo {pages} {061603} (\bibinfo {year} {2016})}\BibitemShut {NoStop}%
\bibitem [{\citenamefont {Kumar}\ \emph {et~al.}(2016)\citenamefont {Kumar}, \citenamefont {Anderson}, \citenamefont {Phillips}, \citenamefont {Eckel}, \citenamefont {Campbell},\ and\ \citenamefont {Stringari}}]{Kumar2016}%
  \BibitemOpen
  \bibfield  {author} {\bibinfo {author} {\bibfnamefont {A.}~\bibnamefont {Kumar}}, \bibinfo {author} {\bibfnamefont {N.}~\bibnamefont {Anderson}}, \bibinfo {author} {\bibfnamefont {W.~D.}\ \bibnamefont {Phillips}}, \bibinfo {author} {\bibfnamefont {S.}~\bibnamefont {Eckel}}, \bibinfo {author} {\bibfnamefont {G.~K.}\ \bibnamefont {Campbell}},\ and\ \bibinfo {author} {\bibfnamefont {S.}~\bibnamefont {Stringari}},\ }\bibfield  {title} {\bibinfo {title} {Minimally destructive, doppler measurement of a quantized flow in a ring-shaped bose--einstein condensate},\ }\href {https://doi.org/10.1088/1367-2630/18/2/025001} {\bibfield  {journal} {\bibinfo  {journal} {New Journal of Physics}\ }\textbf {\bibinfo {volume} {18}},\ \bibinfo {pages} {025001} (\bibinfo {year} {2016})}\BibitemShut {NoStop}%
\bibitem [{\citenamefont {Chauveau}\ \emph {et~al.}(2023)\citenamefont {Chauveau}, \citenamefont {Maury}, \citenamefont {Rabec}, \citenamefont {Heintze}, \citenamefont {Brochier}, \citenamefont {Nascimbene}, \citenamefont {Dalibard}, \citenamefont {Beugnon}, \citenamefont {Roccuzzo},\ and\ \citenamefont {Stringari}}]{Chauveau2023}%
  \BibitemOpen
  \bibfield  {author} {\bibinfo {author} {\bibfnamefont {G.}~\bibnamefont {Chauveau}}, \bibinfo {author} {\bibfnamefont {C.}~\bibnamefont {Maury}}, \bibinfo {author} {\bibfnamefont {F.}~\bibnamefont {Rabec}}, \bibinfo {author} {\bibfnamefont {C.}~\bibnamefont {Heintze}}, \bibinfo {author} {\bibfnamefont {G.}~\bibnamefont {Brochier}}, \bibinfo {author} {\bibfnamefont {S.}~\bibnamefont {Nascimbene}}, \bibinfo {author} {\bibfnamefont {J.}~\bibnamefont {Dalibard}}, \bibinfo {author} {\bibfnamefont {J.}~\bibnamefont {Beugnon}}, \bibinfo {author} {\bibfnamefont {S.~M.}\ \bibnamefont {Roccuzzo}},\ and\ \bibinfo {author} {\bibfnamefont {S.}~\bibnamefont {Stringari}},\ }\bibfield  {title} {\bibinfo {title} {Superfluid fraction in an interacting spatially modulated bose-einstein condensate},\ }\href {https://doi.org/10.1103/PhysRevLett.130.226003} {\bibfield  {journal} {\bibinfo  {journal} {Phys. Rev. Lett.}\ }\textbf {\bibinfo {volume} {130}},\ \bibinfo {pages} {226003} (\bibinfo {year} {2023})}\BibitemShut {NoStop}%
\bibitem [{\citenamefont {Pomeau}\ and\ \citenamefont {Rica}(1994)}]{Pomeau94}%
  \BibitemOpen
  \bibfield  {author} {\bibinfo {author} {\bibfnamefont {Y.}~\bibnamefont {Pomeau}}\ and\ \bibinfo {author} {\bibfnamefont {S.}~\bibnamefont {Rica}},\ }\bibfield  {title} {\bibinfo {title} {Dynamics of a model of supersolid},\ }\href {https://doi.org/10.1103/PhysRevLett.72.2426} {\bibfield  {journal} {\bibinfo  {journal} {Phys. Rev. Lett.}\ }\textbf {\bibinfo {volume} {72}},\ \bibinfo {pages} {2426} (\bibinfo {year} {1994})}\BibitemShut {NoStop}%
\bibitem [{\citenamefont {Josserand}\ \emph {et~al.}(2007{\natexlab{b}})\citenamefont {Josserand}, \citenamefont {Pomeau},\ and\ \citenamefont {Rica}}]{josserand2007}%
  \BibitemOpen
  \bibfield  {author} {\bibinfo {author} {\bibfnamefont {C.}~\bibnamefont {Josserand}}, \bibinfo {author} {\bibfnamefont {Y.}~\bibnamefont {Pomeau}},\ and\ \bibinfo {author} {\bibfnamefont {S.}~\bibnamefont {Rica}},\ }\bibfield  {title} {\bibinfo {title} {Patterns and supersolids},\ }\href {https://doi.org/10.1140/epjst/e2007-00168-9} {\bibfield  {journal} {\bibinfo  {journal} {The European Physical Journal Special Topics}\ }\textbf {\bibinfo {volume} {146}},\ \bibinfo {pages} {47} (\bibinfo {year} {2007}{\natexlab{b}})}\BibitemShut {NoStop}%
\bibitem [{\citenamefont {Chaikin}\ and\ \citenamefont {Lubensky}(1995)}]{chaikin_lubensky_1995}%
  \BibitemOpen
  \bibfield  {author} {\bibinfo {author} {\bibfnamefont {P.~M.}\ \bibnamefont {Chaikin}}\ and\ \bibinfo {author} {\bibfnamefont {T.~C.}\ \bibnamefont {Lubensky}},\ }\href {https://doi.org/10.1017/CBO9780511813467} {\emph {\bibinfo {title} {Principles of Condensed Matter Physics}}}\ (\bibinfo  {publisher} {Cambridge University Press},\ \bibinfo {year} {1995})\BibitemShut {NoStop}%
\bibitem [{\citenamefont {Landau}(1941)}]{landau41}%
  \BibitemOpen
  \bibfield  {author} {\bibinfo {author} {\bibfnamefont {L.}~\bibnamefont {Landau}},\ }\bibfield  {title} {\bibinfo {title} {Theory of the superfluidity of helium ii},\ }\href {https://doi.org/10.1103/PhysRev.60.356} {\bibfield  {journal} {\bibinfo  {journal} {Phys. Rev.}\ }\textbf {\bibinfo {volume} {60}},\ \bibinfo {pages} {356} (\bibinfo {year} {1941})}\BibitemShut {NoStop}%
\bibitem [{\citenamefont {Pitaevskii}\ and\ \citenamefont {Stringari}(2016)}]{BecBook2016}%
  \BibitemOpen
  \bibfield  {author} {\bibinfo {author} {\bibfnamefont {L.}~\bibnamefont {Pitaevskii}}\ and\ \bibinfo {author} {\bibfnamefont {S.}~\bibnamefont {Stringari}},\ }\href@noop {} {\emph {\bibinfo {title} {Bose-Einstein condensation and superfluidity}}}\ (\bibinfo  {publisher} {Oxford University Press},\ \bibinfo {year} {2016})\BibitemShut {NoStop}%
\bibitem [{\citenamefont {Martin}\ \emph {et~al.}(1972)\citenamefont {Martin}, \citenamefont {Parodi},\ and\ \citenamefont {Pershan}}]{Martin1972}%
  \BibitemOpen
  \bibfield  {author} {\bibinfo {author} {\bibfnamefont {P.~C.}\ \bibnamefont {Martin}}, \bibinfo {author} {\bibfnamefont {O.}~\bibnamefont {Parodi}},\ and\ \bibinfo {author} {\bibfnamefont {P.~S.}\ \bibnamefont {Pershan}},\ }\bibfield  {title} {\bibinfo {title} {Unified hydrodynamic theory for crystals, liquid crystals, and normal fluids},\ }\href {https://doi.org/10.1103/PhysRevA.6.2401} {\bibfield  {journal} {\bibinfo  {journal} {Phys. Rev. A}\ }\textbf {\bibinfo {volume} {6}},\ \bibinfo {pages} {2401} (\bibinfo {year} {1972})}\BibitemShut {NoStop}%
\bibitem [{\citenamefont {W.F.Vinen}(1970)}]{Vinen70}%
  \BibitemOpen
  \bibfield  {author} {\bibinfo {author} {\bibnamefont {W.F.Vinen}},\ }\href@noop {} {\emph {\bibinfo {title} {Physics of Quantum Fluids, ed. by R. Kubo and F. Takano}}}\ (\bibinfo  {publisher} {Syokabo Pub- lishing Company, Tokyo},\ \bibinfo {year} {1970})\BibitemShut {NoStop}%
\bibitem [{\citenamefont {Hu}\ \emph {et~al.}(2010)\citenamefont {Hu}, \citenamefont {Taylor}, \citenamefont {Liu}, \citenamefont {Stringari},\ and\ \citenamefont {Griffin}}]{Hu2010}%
  \BibitemOpen
  \bibfield  {author} {\bibinfo {author} {\bibfnamefont {H.}~\bibnamefont {Hu}}, \bibinfo {author} {\bibfnamefont {E.}~\bibnamefont {Taylor}}, \bibinfo {author} {\bibfnamefont {X.-J.}\ \bibnamefont {Liu}}, \bibinfo {author} {\bibfnamefont {S.}~\bibnamefont {Stringari}},\ and\ \bibinfo {author} {\bibfnamefont {A.}~\bibnamefont {Griffin}},\ }\bibfield  {title} {\bibinfo {title} {Second sound and the density response function in uniform superfluid atomic gases},\ }\href {https://doi.org/10.1088/1367-2630/12/4/043040} {\bibfield  {journal} {\bibinfo  {journal} {New Journal of Physics}\ }\textbf {\bibinfo {volume} {12}},\ \bibinfo {pages} {043040} (\bibinfo {year} {2010})}\BibitemShut {NoStop}%
\bibitem [{\citenamefont {Santos}\ \emph {et~al.}(2003)\citenamefont {Santos}, \citenamefont {Shlyapnikov},\ and\ \citenamefont {Lewenstein}}]{SantosRoton}%
  \BibitemOpen
  \bibfield  {author} {\bibinfo {author} {\bibfnamefont {L.}~\bibnamefont {Santos}}, \bibinfo {author} {\bibfnamefont {G.~V.}\ \bibnamefont {Shlyapnikov}},\ and\ \bibinfo {author} {\bibfnamefont {M.}~\bibnamefont {Lewenstein}},\ }\bibfield  {title} {\bibinfo {title} {Roton-maxon spectrum and stability of trapped dipolar bose-einstein condensates},\ }\href {https://doi.org/10.1103/PhysRevLett.90.250403} {\bibfield  {journal} {\bibinfo  {journal} {Phys. Rev. Lett.}\ }\textbf {\bibinfo {volume} {90}},\ \bibinfo {pages} {250403} (\bibinfo {year} {2003})}\BibitemShut {NoStop}%
\bibitem [{\citenamefont {Petter}\ \emph {et~al.}(2019)\citenamefont {Petter}, \citenamefont {Natale}, \citenamefont {van Bijnen}, \citenamefont {Patscheider}, \citenamefont {Mark}, \citenamefont {Chomaz},\ and\ \citenamefont {Ferlaino}}]{petter-roton2019}%
  \BibitemOpen
  \bibfield  {author} {\bibinfo {author} {\bibfnamefont {D.}~\bibnamefont {Petter}}, \bibinfo {author} {\bibfnamefont {G.}~\bibnamefont {Natale}}, \bibinfo {author} {\bibfnamefont {R.~M.~W.}\ \bibnamefont {van Bijnen}}, \bibinfo {author} {\bibfnamefont {A.}~\bibnamefont {Patscheider}}, \bibinfo {author} {\bibfnamefont {M.~J.}\ \bibnamefont {Mark}}, \bibinfo {author} {\bibfnamefont {L.}~\bibnamefont {Chomaz}},\ and\ \bibinfo {author} {\bibfnamefont {F.}~\bibnamefont {Ferlaino}},\ }\bibfield  {title} {\bibinfo {title} {Probing the roton excitation spectrum of a stable dipolar bose gas},\ }\href {https://doi.org/10.1103/PhysRevLett.122.183401} {\bibfield  {journal} {\bibinfo  {journal} {Phys. Rev. Lett.}\ }\textbf {\bibinfo {volume} {122}},\ \bibinfo {pages} {183401} (\bibinfo {year} {2019})}\BibitemShut {NoStop}%
\bibitem [{Note2()}]{Note2}%
  \BibitemOpen
  \bibinfo {note} {Ring configurations of similar shape, hosting a supersolid dipolar gas with persistent currents, were considered in \cite {tengstrand2021}.}\BibitemShut {Stop}%
\bibitem [{\citenamefont {Tao}\ \emph {et~al.}(2023)\citenamefont {Tao}, \citenamefont {Zhao},\ and\ \citenamefont {Spielman}}]{Tao2023}%
  \BibitemOpen
  \bibfield  {author} {\bibinfo {author} {\bibfnamefont {J.}~\bibnamefont {Tao}}, \bibinfo {author} {\bibfnamefont {M.}~\bibnamefont {Zhao}},\ and\ \bibinfo {author} {\bibfnamefont {I.~B.}\ \bibnamefont {Spielman}},\ }\bibfield  {title} {\bibinfo {title} {Observation of anisotropic superfluid density in an artificial crystal},\ }\href {https://doi.org/10.1103/PhysRevLett.131.163401} {\bibfield  {journal} {\bibinfo  {journal} {Phys. Rev. Lett.}\ }\textbf {\bibinfo {volume} {131}},\ \bibinfo {pages} {163401} (\bibinfo {year} {2023})}\BibitemShut {NoStop}%
\bibitem [{\citenamefont {Smith}\ \emph {et~al.}(2023)\citenamefont {Smith}, \citenamefont {Baillie},\ and\ \citenamefont {Blakie}}]{Smith2023}%
  \BibitemOpen
  \bibfield  {author} {\bibinfo {author} {\bibfnamefont {J.~C.}\ \bibnamefont {Smith}}, \bibinfo {author} {\bibfnamefont {D.}~\bibnamefont {Baillie}},\ and\ \bibinfo {author} {\bibfnamefont {P.~B.}\ \bibnamefont {Blakie}},\ }\bibfield  {title} {\bibinfo {title} {Supersolidity and crystallization of a dipolar bose gas in an infinite tube},\ }\href {https://doi.org/10.1103/PhysRevA.107.033301} {\bibfield  {journal} {\bibinfo  {journal} {Phys. Rev. A}\ }\textbf {\bibinfo {volume} {107}},\ \bibinfo {pages} {033301} (\bibinfo {year} {2023})}\BibitemShut {NoStop}%
\bibitem [{Note3()}]{Note3}%
  \BibitemOpen
  \bibinfo {note} {The energy dependence on angular momentum for such rotating configurations at larger $\Omega $ has recently been explored in \cite {tengstrand2021}.}\BibitemShut {Stop}%
\bibitem [{Note4()}]{Note4}%
  \BibitemOpen
  \bibinfo {note} {The good agreement between Leggett's upper bound and the superfluid fraction calculated in an infinite tube by means of effective mass was recently discussed in \cite {Smith2023}.}\BibitemShut {Stop}%
\bibitem [{\citenamefont {Saito}\ \emph {et~al.}(2009)\citenamefont {Saito}, \citenamefont {Kawaguchi},\ and\ \citenamefont {Ueda}}]{Ueda2009}%
  \BibitemOpen
  \bibfield  {author} {\bibinfo {author} {\bibfnamefont {H.}~\bibnamefont {Saito}}, \bibinfo {author} {\bibfnamefont {Y.}~\bibnamefont {Kawaguchi}},\ and\ \bibinfo {author} {\bibfnamefont {M.}~\bibnamefont {Ueda}},\ }\bibfield  {title} {\bibinfo {title} {Ferrofluidity in a two-component dipolar bose-einstein condensate},\ }\href {https://doi.org/10.1103/PhysRevLett.102.230403} {\bibfield  {journal} {\bibinfo  {journal} {Phys. Rev. Lett.}\ }\textbf {\bibinfo {volume} {102}},\ \bibinfo {pages} {230403} (\bibinfo {year} {2009})}\BibitemShut {NoStop}%
\bibitem [{\citenamefont {Bland}\ \emph {et~al.}(2022)\citenamefont {Bland}, \citenamefont {Poli}, \citenamefont {Ardila}, \citenamefont {Santos}, \citenamefont {Ferlaino},\ and\ \citenamefont {Bisset}}]{Bland2022-dipmix}%
  \BibitemOpen
  \bibfield  {author} {\bibinfo {author} {\bibfnamefont {T.}~\bibnamefont {Bland}}, \bibinfo {author} {\bibfnamefont {E.}~\bibnamefont {Poli}}, \bibinfo {author} {\bibfnamefont {L.~A. P.~n.}\ \bibnamefont {Ardila}}, \bibinfo {author} {\bibfnamefont {L.}~\bibnamefont {Santos}}, \bibinfo {author} {\bibfnamefont {F.}~\bibnamefont {Ferlaino}},\ and\ \bibinfo {author} {\bibfnamefont {R.~N.}\ \bibnamefont {Bisset}},\ }\bibfield  {title} {\bibinfo {title} {Alternating-domain supersolids in binary dipolar condensates},\ }\href {https://doi.org/10.1103/PhysRevA.106.053322} {\bibfield  {journal} {\bibinfo  {journal} {Phys. Rev. A}\ }\textbf {\bibinfo {volume} {106}},\ \bibinfo {pages} {053322} (\bibinfo {year} {2022})}\BibitemShut {NoStop}%
\bibitem [{\citenamefont {Li}\ \emph {et~al.}(2022)\citenamefont {Li}, \citenamefont {Le},\ and\ \citenamefont {Saito}}]{Li2022-dipmix}%
  \BibitemOpen
  \bibfield  {author} {\bibinfo {author} {\bibfnamefont {S.}~\bibnamefont {Li}}, \bibinfo {author} {\bibfnamefont {U.~N.}\ \bibnamefont {Le}},\ and\ \bibinfo {author} {\bibfnamefont {H.}~\bibnamefont {Saito}},\ }\bibfield  {title} {\bibinfo {title} {Long-lifetime supersolid in a two-component dipolar bose-einstein condensate},\ }\href {https://doi.org/10.1103/PhysRevA.105.L061302} {\bibfield  {journal} {\bibinfo  {journal} {Phys. Rev. A}\ }\textbf {\bibinfo {volume} {105}},\ \bibinfo {pages} {L061302} (\bibinfo {year} {2022})}\BibitemShut {NoStop}%
\bibitem [{\citenamefont {Scheiermann}\ \emph {et~al.}(2023)\citenamefont {Scheiermann}, \citenamefont {Ardila}, \citenamefont {Bland}, \citenamefont {Bisset},\ and\ \citenamefont {Santos}}]{scheiermann2023}%
  \BibitemOpen
  \bibfield  {author} {\bibinfo {author} {\bibfnamefont {D.}~\bibnamefont {Scheiermann}}, \bibinfo {author} {\bibfnamefont {L.~A. P.~n.}\ \bibnamefont {Ardila}}, \bibinfo {author} {\bibfnamefont {T.}~\bibnamefont {Bland}}, \bibinfo {author} {\bibfnamefont {R.~N.}\ \bibnamefont {Bisset}},\ and\ \bibinfo {author} {\bibfnamefont {L.}~\bibnamefont {Santos}},\ }\bibfield  {title} {\bibinfo {title} {Catalyzation of supersolidity in binary dipolar condensates},\ }\href {https://doi.org/10.1103/PhysRevA.107.L021302} {\bibfield  {journal} {\bibinfo  {journal} {Phys. Rev. A}\ }\textbf {\bibinfo {volume} {107}},\ \bibinfo {pages} {L021302} (\bibinfo {year} {2023})}\BibitemShut {NoStop}%
\bibitem [{\citenamefont {Kirkby}\ \emph {et~al.}(2023)\citenamefont {Kirkby}, \citenamefont {Lee}, \citenamefont {Baillie}, \citenamefont {Bland}, \citenamefont {Ferlaino}, \citenamefont {Blakie},\ and\ \citenamefont {Bisset}}]{kirkby2023excitations}%
  \BibitemOpen
  \bibfield  {author} {\bibinfo {author} {\bibfnamefont {W.}~\bibnamefont {Kirkby}}, \bibinfo {author} {\bibfnamefont {A.-C.}\ \bibnamefont {Lee}}, \bibinfo {author} {\bibfnamefont {D.}~\bibnamefont {Baillie}}, \bibinfo {author} {\bibfnamefont {T.}~\bibnamefont {Bland}}, \bibinfo {author} {\bibfnamefont {F.}~\bibnamefont {Ferlaino}}, \bibinfo {author} {\bibfnamefont {P.~B.}\ \bibnamefont {Blakie}},\ and\ \bibinfo {author} {\bibfnamefont {R.~N.}\ \bibnamefont {Bisset}},\ }\href@noop {} {\bibinfo {title} {Excitations of a binary supersolid}} (\bibinfo {year} {2023}),\ \Eprint {https://arxiv.org/abs/2312.03390} {arXiv:2312.03390 [cond-mat.quant-gas]} \BibitemShut {NoStop}%
\bibitem [{\citenamefont {Rapaport}\ and\ \citenamefont {Chen}(2007)}]{Rapaport_2007}%
  \BibitemOpen
  \bibfield  {author} {\bibinfo {author} {\bibfnamefont {R.}~\bibnamefont {Rapaport}}\ and\ \bibinfo {author} {\bibfnamefont {G.}~\bibnamefont {Chen}},\ }\bibfield  {title} {\bibinfo {title} {Experimental methods and analysis of cold and dense dipolar exciton fluids},\ }\href {https://doi.org/10.1088/0953-8984/19/29/295207} {\bibfield  {journal} {\bibinfo  {journal} {Journal of Physics: Condensed Matter}\ }\textbf {\bibinfo {volume} {19}},\ \bibinfo {pages} {295207} (\bibinfo {year} {2007})}\BibitemShut {NoStop}%
\bibitem [{\citenamefont {Combescot}\ \emph {et~al.}(2017)\citenamefont {Combescot}, \citenamefont {Combescot},\ and\ \citenamefont {Dubin}}]{Combescot_rev_2017}%
  \BibitemOpen
  \bibfield  {author} {\bibinfo {author} {\bibfnamefont {M.}~\bibnamefont {Combescot}}, \bibinfo {author} {\bibfnamefont {R.}~\bibnamefont {Combescot}},\ and\ \bibinfo {author} {\bibfnamefont {F.}~\bibnamefont {Dubin}},\ }\bibfield  {title} {\bibinfo {title} {Bose--einstein condensation and indirect excitons: a review},\ }\href {https://doi.org/10.1088/1361-6633/aa50e3} {\bibfield  {journal} {\bibinfo  {journal} {Reports on Progress in Physics}\ }\textbf {\bibinfo {volume} {80}},\ \bibinfo {pages} {066501} (\bibinfo {year} {2017})}\BibitemShut {NoStop}%
\bibitem [{\citenamefont {Butov}\ \emph {et~al.}(2002)\citenamefont {Butov}, \citenamefont {Gossard},\ and\ \citenamefont {Chemla}}]{ButovNature2002}%
  \BibitemOpen
  \bibfield  {author} {\bibinfo {author} {\bibfnamefont {L.~V.}\ \bibnamefont {Butov}}, \bibinfo {author} {\bibfnamefont {A.~C.}\ \bibnamefont {Gossard}},\ and\ \bibinfo {author} {\bibfnamefont {D.~S.}\ \bibnamefont {Chemla}},\ }\bibfield  {title} {\bibinfo {title} {Macroscopically ordered state in an exciton system},\ }\href {https://doi.org/10.1038/nature00943} {\bibfield  {journal} {\bibinfo  {journal} {Nature}\ }\textbf {\bibinfo {volume} {418}},\ \bibinfo {pages} {751} (\bibinfo {year} {2002})}\BibitemShut {NoStop}%
\bibitem [{\citenamefont {Andreev}(2017)}]{Andreev2017}%
  \BibitemOpen
  \bibfield  {author} {\bibinfo {author} {\bibfnamefont {S.~V.}\ \bibnamefont {Andreev}},\ }\bibfield  {title} {\bibinfo {title} {Fragmented-condensate solid of dipolar excitons},\ }\href {https://doi.org/10.1103/PhysRevB.95.184519} {\bibfield  {journal} {\bibinfo  {journal} {Phys. Rev. B}\ }\textbf {\bibinfo {volume} {95}},\ \bibinfo {pages} {184519} (\bibinfo {year} {2017})}\BibitemShut {NoStop}%
\bibitem [{\citenamefont {Tengstrand}\ \emph {et~al.}(2021)\citenamefont {Tengstrand}, \citenamefont {Boholm}, \citenamefont {Sachdeva}, \citenamefont {Bengtsson},\ and\ \citenamefont {Reimann}}]{tengstrand2021}%
  \BibitemOpen
  \bibfield  {author} {\bibinfo {author} {\bibfnamefont {M.~N.}\ \bibnamefont {Tengstrand}}, \bibinfo {author} {\bibfnamefont {D.}~\bibnamefont {Boholm}}, \bibinfo {author} {\bibfnamefont {R.}~\bibnamefont {Sachdeva}}, \bibinfo {author} {\bibfnamefont {J.}~\bibnamefont {Bengtsson}},\ and\ \bibinfo {author} {\bibfnamefont {S.~M.}\ \bibnamefont {Reimann}},\ }\bibfield  {title} {\bibinfo {title} {Persistent currents in toroidal dipolar supersolids},\ }\href {https://doi.org/10.1103/PhysRevA.103.013313} {\bibfield  {journal} {\bibinfo  {journal} {Phys. Rev. A}\ }\textbf {\bibinfo {volume} {103}},\ \bibinfo {pages} {013313} (\bibinfo {year} {2021})}\BibitemShut {NoStop}%
\end{thebibliography}%

\end{document}